\documentclass[prd,aps,floatfix,preprintnumbers,twocolumn,%
               superscriptaddress,showpacs,nofootinbib,byrevtex]{revtex4}
\usepackage{graphicx}
\usepackage{longtable}
\begin{document}

\newcommand{\figscale}{0.55}
\date{\today}

\title{Bulk first-order phase transition in three-flavor lattice QCD with
  $O(a)$-improved Wilson fermion action at zero temperature}

\newcommand{\Tsukuba}%
{Graduate School of Pure and Applied Sciences, University of Tsukuba, 
 Tsukuba, Ibaraki 305-8571, Japan}

\newcommand{\RCCP}%
{Center for Computational Science, University of Tsukuba,
 Tsukuba, Ibaraki 305-8577, Japan}

\newcommand{\ICRR}%
{Institute for Cosmic Ray Research, University of Tokyo, 
 Kashiwa, Chiba 277-8582, Japan}

\newcommand{\Hiroshima}%
{Department of Physics, Hiroshima University,
 Higashi-Hiroshima, Hiroshima 739-8526, Japan}

\newcommand{\KEK}%
{High Energy Accelerator Research Organization (KEK), 
 Tsukuba, Ibaraki 305-0801, Japan}

\newcommand{\YITP}%
{Yukawa Institute for Theoretical Physics, Kyoto University, 
 Kyoto 606-8502, Japan}

\newcommand{\RBRC}%
{RIKEN BNL Research Center, Brookhaven National Laboratory, 
 Upton, NY 11973, USA}

\author{S.~Aoki}
\affiliation{\Tsukuba}

\author{M.~Fukugita}
\affiliation{\ICRR}

\author{S.~Hashimoto}
\affiliation{\KEK}

\author{K-I.~Ishikawa}
\affiliation{\Hiroshima}

\author{N.~Ishizuka}
\affiliation{\Tsukuba}
\affiliation{\RCCP}

\author{Y.~Iwasaki}
\affiliation{\Tsukuba}

\author{K.~Kanaya}
\affiliation{\Tsukuba}
\affiliation{\RCCP}

\author{T.~Kaneko}
\affiliation{\KEK}

\author{Y.~Kuramashi}
\affiliation{\Tsukuba}
\affiliation{\RCCP}

\author{M.~Okawa}
\affiliation{\Hiroshima}

\author{N.~Tsutsui}
\affiliation{\KEK}

\author{A.~Ukawa}
\affiliation{\Tsukuba}
\affiliation{\RCCP}

\author{N.~Yamada}
\affiliation{\RBRC}

\author{T.~Yoshi\'{e}}
\affiliation{\Tsukuba}
\affiliation{\RCCP}

\collaboration{JLQCD Collaboration}
\noaffiliation

\preprint{KEK-CP-151}
\preprint{UTCCS-P-3}
\preprint{UTHEP-489}
\preprint{HUPD-0401}

\begin{abstract}
  Three-flavor QCD simulation with the $O(a)$-improved 
  Wilson fermion action is made employing an exact fermion 
  algorithm developed for odd number of quark flavors. 
  For the plaquette gauge action, an unexpected first-order 
  phase transition is found in the strong coupling regime  
  ($\beta\lesssim$ 5.0) at relatively heavy quark masses 
  ($m_{\mathrm{PS}}/m_{\mathrm{V}}\sim$ 0.74--0.87). 
  Strong metastability persists on a large lattice of 
  size $12^3\times 32$, which indicates that the transition 
  has a bulk nature. 
  The phase gap becomes smaller toward weaker couplings and 
  vanishes at $\beta\simeq 5.0$, which corresponds to  
  a lattice spacing $a\simeq$ 0.1~fm. 
  The phase transition is not found if the improved gauge 
  actions are employed. 
  Our results imply that realistic simulations of QCD with three 
  flavors of dynamical Wilson-type fermions at lattice 
  spacings in the range $a=$ 0.1--0.2~fm require use of improved gauge actions.
  Possible origins of the phase transition is discussed.   
\end{abstract}

\maketitle

\section{Introduction}
Realistic simulations of the strong interaction through 
lattice QCD require dynamical treatment of up, down and strange 
quarks incorporating their pair creation and annihilation effects in the vacuum. 
While simulations with the dynamical up and down quarks have now become
routine (for recent studies, see 
Refs.~\cite{Eicker:1998sy,Allton:1998gi,AliKhan:2001tx,Aoki:2002uc}),
adding a dynamical strange quark is still in the development stage. 
This is primarily because no exact algorithm to treat odd
number of flavors was known until recently.
In fact with the conventional Hybrid Monte Carlo (HMC) algorithm~\cite{Duane:de}, 
the number of flavors is limited to even.
The $R$ algorithm~\cite{Gottlieb:mq} can be used for any
number of flavors, as applied in recent three-flavor 
simulations with the staggered quark 
action~\cite{Bernard:2001av,Davies:2003ik}.  
However, the results are subject to systematic
errors due to a finite step size in the molecular dynamics
evolution, and this has to be controlled by taking the 
limit of zero step size which is quite computer time consuming.

Recently, however, several exact algorithms 
for an arbitrary number of flavors have been proposed 
both for the Wilson-type~\cite{Takaishi:2001um,Aoki:2001pt} 
and the staggered-type~\cite{Aoki:2002xi} fermion actions.  
They have been shown to work with realistic lattice volumes 
without much increase of computational cost compared to 
HMC~\cite{Aoki:2001pt,Aoki:2002xi}. 
Thus, a practical barrier to
perform realistic three-flavor QCD simulations has been eliminated.

The next step toward realistic simulations is to explore the parameter space 
of the three-flavor lattice QCD to ensure that the system is free from 
lattice artifacts over a range of lattice spacing $a$ and quark mass $m_q$ 
adequate for phenomenology. 
In practice this means finding a region corresponding to 
$a\simeq$ 0.1--0.2~fm and $m_q= (1/3\sim 1)\times m_s$, with $m_s$ the
physical strange quark mass.
 
For the Wilson-type fermion action, there is a related important 
theoretical issue to settle.  For even number of flavors, 
evidence has been accumulated 
over the years that vanishing of pion mass at a critical
hopping parameter $\kappa_c(\beta)$ ($\beta=6/g^2$ with $g$ the gauge 
coupling ) is due to spontaneous breaking of flavor-parity rotational
symmetry~\cite{Aoki:1984,Aoki:1996}.
Whether this understanding holds for odd number of flavors still 
needs confirmation.  
It is therefore mandatory to cover the entire parameter
space of the lattice action for three-flavor QCD.

In this work we numerically explore the parameter space 
$(\beta,\kappa)$ of three-flavor lattice QCD with the 
$O(a)$-improved Wilson fermion action~\cite{Sheikholeslami:1985ij}. 
The three quark flavors are assumed to be degenerate in mass.   
For  the gauge action, we test both the plaquette and 
renormalization group (RG) improved gauge actions. 
The $O(a)$-improvement coefficient $c_{\mathrm{SW}}$
is fixed to the one-loop perturbative value as 
fully non-perturbative values for the relevant gauge actions 
were not available until after the present work 
was well in progress~\cite{Aoki:2002vh,Ishikawa:2003ri,Yamada:2004ja}.

For the standard plaquette gauge action, 
we unexpectedly find~\cite{Aoki:2001xq} 
a clear evidence of the existence of an ordered phase 
for large $\kappa$ separated by a first-order phase transition 
from the disordered phase at smaller $\kappa$.  The transition persists 
for large volumes, and hence it is a bulk phase transition. 
Contrary to the disordered phase,  the ordered phase does not 
exhibits the standard features of the confining phase.  
For example, 
the pseudo-scalar-to-vector meson mass ratio is similar in the 
disordered and ordered phases at around 0.8, but 
the lattice spacing is unnaturally large in the ordered phase. 
The first-order phase transition is observed in the strong coupling regime. 
The gap of physical quantities across the transition
diminishes toward weaker couplings, and 
appears to vanish at $\beta\simeq 5.0$, which corresponds 
to a lattice spacing $a\simeq$ 0.1~fm.  

These results indicate that there is a parameter region 
which is not smoothly connected to continuum three-flavor QCD. 
In other words, the continuum three-flavor QCD can be 
approached only if one uses lattices much finer than $a\simeq$ 
0.1~fm, if one employs the plaquette gauge action. 

We find that the metastability signals disappear 
if one employs improved gauge actions such as the
renormalization group (RG) improved action~\cite{Iwasaki:1983ck}
or $O(a^2)$-improved L\"uscher-Weisz action~\cite{Luscher:1984xn}. 
With these actions, the continuum extrapolation should be possible 
from a conventional range of lattice spacings $a\sim 0.1--0.2$~fm.  

The finding of a first-order phase transition for the plaquette gauge 
action is quite unexpected.  A large number of simulations 
carried out in the past for the quenched and two-flavor cases were 
consistent with 
the expectation that the disordered ({\it i.e.,} confining ) phase extends over 
$0\leq \kappa\leq \kappa_c$ for any value of $\beta$.  
Recently, however, while making a study of twisted mass QCD, 
Farchioni {\it et al.} reported in a two-flavor simulation 
with unimproved Wilson quark action and plaquette gauge 
action~\cite{Farchioni:2004us} that there exists a first-order phase transition 
at $\beta=5.2$.  They suggested that this phase transition can be understood 
as the alternative to the parity-flavor broken phase which was pointed out by  
Sharpe and Singleton~\cite{Sharpe:1998xm}.

It is possible that their finding and ours have a common origin.  
Another possible explanation for our first-order transition is that 
it is related to the 
first-order phase transition encountered in pure $SU(3)$ gauge theory 
in the extended coupling space $(\beta,\beta_A)$, where $\beta_A$
characterizes the strength of adjoint 
representation~\cite{Greensite:hw,Bhanot:1981pj,Heller:1995bz}. 
Further work is needed for clarification of the origin of the first order-phase 
transition both for the two and three flavor cases.  

The rest of this paper is organized as follows.
In Section~\ref{sec:Lattice_action_and_algorithm} we
introduce the lattice actions and simulation algorithms we
employed. 
Section~\ref{sec:Phase_structure} describes our study of
phase structure of three-flavor lattice QCD with the plaquette 
gauge action, where the
presence of the metastable states is discussed in detail.
The phase structure analysis for the case of improved gauge actions 
is discussed in \ref{sec:improved_actions}.  
In Section~\ref{sec:Discussions} we discuss possible origin 
of the first-order transition and related phenomena encountered in 
past studies.  A conclusion is given in \ref{sec:Conclusions} 
where we also discuss the possibility of realistic simulations of
three-flavor QCD.  The first report of the present work was briefly 
made in Ref.~\cite{Aoki:2001xq}

\section{Lattice action and algorithm}
\label{sec:Lattice_action_and_algorithm}

The partition function we study is defined by
\begin{equation}
  {\cal Z}=\int\!\!{\cal D}U\,
  (\det[D_{ud}])^2(\det[D_s])e^{-S_{g}(U)}.
  \label{eq:pertition_function}
\end{equation}
Here $S_g(U)$ is the gluon action given by  
\begin{equation}
  S_g(U)=
  {\beta \over 6}\left[c_0 \sum W_{1 \times 1} + c_1\sum W_{1 \times 2}\right],
  \label{eq:gluon_actions}
\end{equation}
where $W_{1 \times 1}$ and $W_{1 \times 2}$
are the plaquette and rectangular Wilson loops,
respectively.
The summations are taken over all possible plaquettes and
rectangles on the lattice.
The coefficients $c_0$ and $c_1$ are determined as
$c_0=1-8c_1$, with
$c_1$ = 0, $-0.331$, or $-1/(12\langle P\rangle^{1/2})$ for
the standard Wilson action, the RG-improved action~\cite{Iwasaki:1983ck}, 
and 
the $O(a^2)$-improved (L\"uscher-Weisz (LW)) action~\cite{Luscher:1984xn}, 
respectively, and $\langle P\rangle$ is the plaquette average 
introduced for a mean field improvement.

The fermionic determinant $(\det[D_{ud}])^2$ represents the
contribution of degenerate up ($u$) and down ($d$) quarks
whereas the strange ($s$) quark effect is given by $(\det[D_s])$. 
In this work we mainly consider the $O(a)$-improved
Wilson-Dirac operator 
$D_q=1+M_q+T_q$ ($q=u, d, s$) with $M_q$ the usual hopping
term including the hopping parameter $\kappa_q$ 
and $T_q$ the $O(a)$-improvement 
SW term~\cite{Sheikholeslami:1985ij}. 
The explicit form of $T_q$ is given by
\begin{equation}
  T_q = 
  -\frac{1}{2}
  c_{\mathrm{SW}} \kappa_q \sigma_{\mu \nu} {\cal F}_{\mu \nu},
  \label{eq:SWterm}
\end{equation} 
with ${\cal F}_{\mu \nu}$ the clover-leaf-type field
strength on the lattice.
For the coefficient $c_{\mathrm{SW}}$ we employ the value determined by  
tadpole improved one-loop perturbation theory as
\begin{equation}
  c_{\mathrm{SW}}={1 \over \langle P\rangle^{3/4}} 
  \left(
    1+c^{(1)}_{\mathrm{SW}}
    \frac{6/\beta}{\langle P\rangle}
  \right),
  \label{eq:c_sw_plaqutte}
\end{equation}
where 
$c^{(1)}_{\mathrm{SW}}=0.0159$~\cite{Wohlert:1987rf,Naik:1993ux,Luscher:1996vw}
for the standard Wilson gauge action.
For improved gauge actions it becomes
$c^{(1)}_{\mathrm{SW}}=0.008$ (RG) or $0.013$ (LW)~\cite{Aoki:1998qd}.
The average plaquette $\langle P\rangle$ is calculated in
pure gauge theory with the same value of $\beta$.

We employ an exact HMC-type algorithm for three-flavor QCD developed in 
Ref.~\cite{Aoki:2001pt}.   
The $u$ and $d$ quark determinant $(\det[D_{ud}])^2$ is estimated 
by the usual pseudo-fermion integral: 
\begin{equation}
  (\det[D_{ud}])^2  =
  \int\!\!{\cal D}\phi_{ud}^{\dag}{\cal D}\phi_{ud}
  \exp\left[-|D_{ud}^{-1}\phi_{ud}|^2\right].
  \label{eq:det[D_u]}
\end{equation}
To represent $(\det[D_s])$ in a similar manner, 
we approximate the inverse of $D_s$ by the 
non-Hermitian Chebyshev polynomial~\cite{Borici:1995am,Borrelli:1996re}
of order $2n$ :
\begin{eqnarray}
  1/D_s \approx P_{2n}(D_s) 
       &\equiv& \sum_{i=0}^{2n}c_i(D_s-1)^i \nonumber\\
       &=&\prod_{k=1}^{n}(D_s-z_{j(k)}^*)(D_s-z_{j(k)}),
  \label{eq:1/D_s}
\end{eqnarray}
with $z_k = 1-\exp(i\ 2\pi k/(2n+1))$~\cite{Alexandrou:1998wv}
and a reordering index $j(k)$.  We then rewrite
\begin{eqnarray}
  (\det[D_s]) &=& \det[D_s P_{2n}(D_s)] \nonumber\\
  &&\times
  \int\!\!{\cal D}\phi_s^{\dag}{\cal D}\phi_s
  \exp\left[-|T_n(D_s)\phi_s|^2\right],
  \label{eq:det[D_s]}
\end{eqnarray}
with 
$T_n(D_s)\equiv\sum_{i=0}^{n} d_{i}(D_s-1)^{i}(=\prod_{k=1}^{n}(D_s-z_{j(k)})$.
Introducing a fictitious momentum $P$ conjugate to the link 
variable $U$, the effective Hamiltonian for the 2+1-flavor
QCD reads 
\begin{equation}
  H=\frac{1}{2}P^2 + S_g(U) + |D_{ud}^{-1}\phi_{ud}|^2  +
  |T_n(D_s)\phi_s|^2. 
  \label{eq:hamiltonian}
\end{equation}
We take account of the correction factor 
$\det[D_s P_{2n}(D_s)]$ by the noisy Metropolis test.
After a trial configuration $U'$ is accepted by the usual
HMC Metropolis test, we make another Metropolis test with
the acceptance probability
$P_{\mathrm{corr}}[U\to U']=\min[1,e^{-dS}]$ 
with 
$dS = |A(U')^{-1}A(U)\chi|^2-|\chi|^2$.
Here $A=[D_s P_{2n}(D_s)]^{1/2}$ and $\chi$ is the
Gaussian noise vector with an unit variance.
For other details of the algorithm, see Ref.~\cite{Aoki:2001pt}.

\section{Phase structure for the plaquette gauge action}
\label{sec:Phase_structure}

\subsection{Thermal cycle analysis}
We study the phase structure of three-flavor lattice QCD for the plaquette 
gauge action ($c_1=0$) assuming flavor degeneracy
$\kappa\equiv\kappa_{ud}=\kappa_s$.
Rapid thermal cycles are performed in the
$(\beta,\kappa)$ parameter space on $4^3 \times 16$ and 
$8^3\times 16$ lattices.
For a fixed value of $\beta$ we start a HMC
simulation at $\kappa=0$ and increase $\kappa$ in unit of
0.001 at every 200 HMC trajectories.
This process is continued until we reach the point at which
we encounter a large violation of HMC energy conservation satisfying $dH>100$.
Then, we reverse the process and decrease $\kappa$ until it
reaches the point sufficiently far away from the turning point. 
This procedure is repeated at a fixed interval over a range of $\beta$.

In this global scan, we do not perform the global
Metropolis test of HMC or the noisy Metropolis test for the
correction factor $\det[D_s P_{2n}(D_s)]$ in
Eq.~(\ref{eq:det[D_s]}).
The order of the Chebyshev polynomial is fixed to $2n=100$ and
the molecular dynamics step size to $d\tau=1/40$ employing 
$\tau=1$ for the length of unit trajectory. 
The stopping criterion of the BiCGStab solver is such that
the residual defined by $|D x-b|/|b|$ becomes smaller than
$10^{-14}$ ($10^{-8}$) for Hamiltonian (force) calculation,
where $D$ is the even-odd preconditioned $O(a)$-improved
Wilson-Dirac operator, $x$ is the solution vector, and $b$
is a source vector. 
The expectation values of observables are measured during the last 100 
trajectories after 100 thermalization trajectories at each $\kappa$
in the cycles.

\begin{figure}[tbp]
  \centering
  \includegraphics[scale=\figscale]{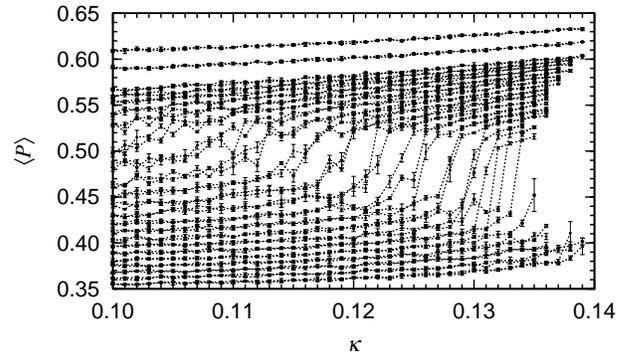}
  \caption{
    Thermal cycles of averaged plaquette $\langle P\rangle$
    on a $4^3 \times 16$ lattice at 
    $\beta$ = 4.6--5.6 with 0.05 steps and 5.8, 6.0 from
    bottom to top.
    Each points are measured on 100 trajectories followed by
    100 thermalization trajectories from its previous
    point. 
  }
\label{fig:thermal_cycle_P_4x8}
\end{figure}

In Figure~\ref{fig:thermal_cycle_P_4x8} we present the plaquette expectation 
value $\langle P\rangle$ during thermal cycles on a $4^3\times 16$ lattice.  
The value of $\beta$ increases from 
$\beta=4.6$ to $5.6$ from bottom to top in units of $0.05$.  We 
observe strong indication of metastability at $4.8\leq \beta\leq 5.1$. 
The system appears to jump from a disordered phase at smaller $\kappa$ to 
an ordered phase at larger $\kappa$. 

\begin{figure}[tbp]
  \centering
  \includegraphics[scale=\figscale]{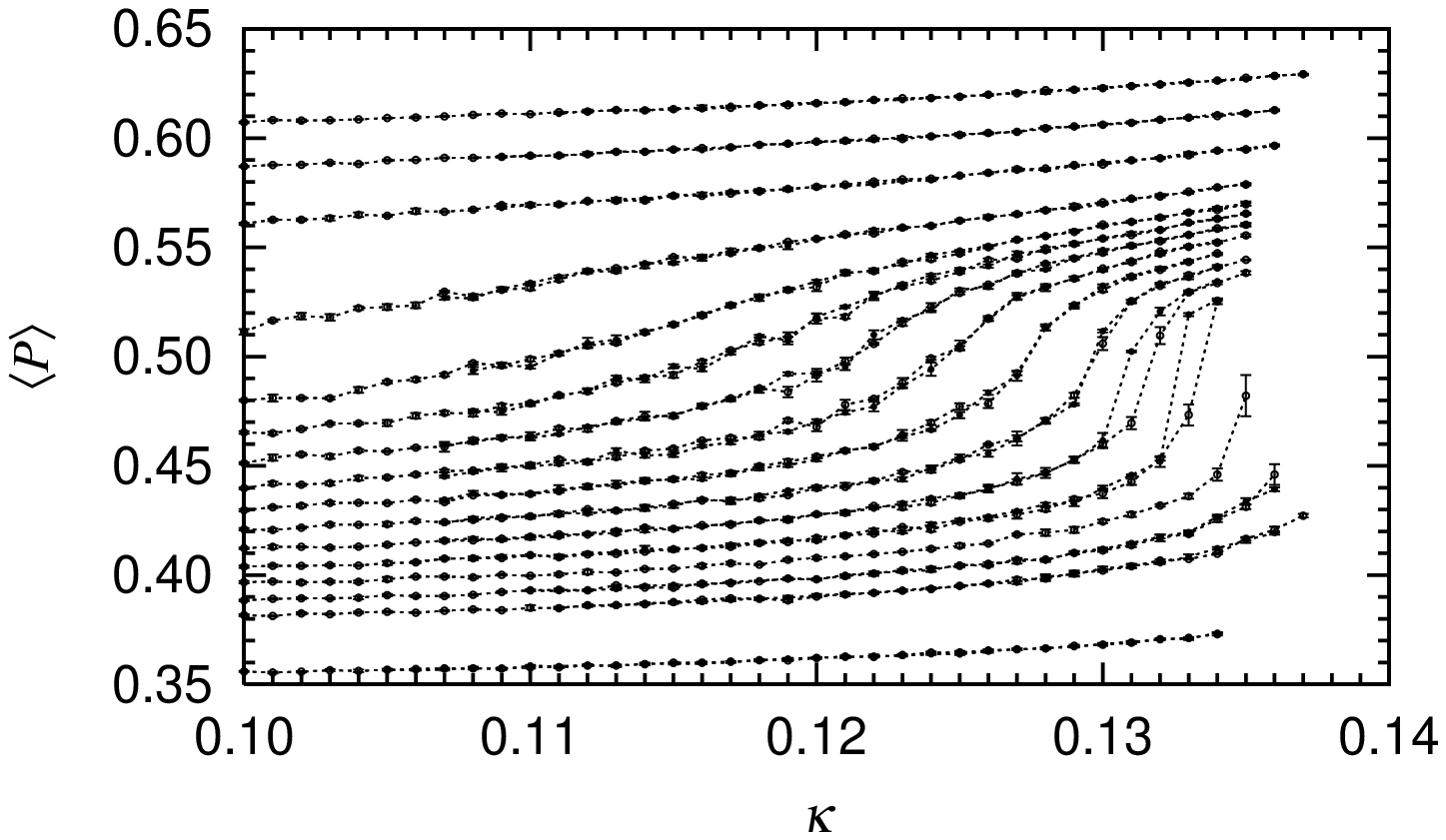}
  \caption{
    Thermal cycles of averaged plaquette $\langle P\rangle$
    on a $8^3 \times 16$ lattice at 
    $\beta$ = 4.6, 4.8--5.3 with 0.05 steps, and 5.4--6.0 with 0.1 steps (from
    bottom to top) for the standard Wilson gauge action.
    Each points are measured on 100 trajectories followed by
    100 thermalization trajectories from its previous
    point.
  }
\label{fig:thermal_cycle_P_8x16}
\end{figure}

\begin{figure}[tbp]
  \centering
  \includegraphics[scale=\figscale]{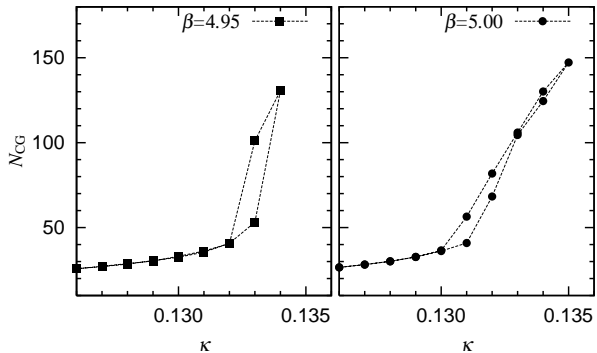}
  \caption{
    Number of BiCGStab steps $N_{\mathrm{CG}}$ required for
    the fermion matrix inversion during the HMC trajectories
    at $\beta$ = 4.95 and 5.0 for the standard Wilson gauge
    action. }
\label{fig:NCG_thermal_cycle_P_8x16}
\end{figure}

On small lattices such as $4^3 \times 16$, the gap and metastability 
might be attributed to a  ``finite-temperature phase transition'' due to 
a small spatial size, corresponding to the thermal first-order transition 
observed for three-flavor QCD with the (unimproved)
Wilson fermion action~\cite{Iwasaki:1996zt}.
To examine this possibility, we show 
$\langle P\rangle$ during the thermal cycles on a spatially larger 
$8^3 \times 16$ lattice in Figure~\ref{fig:thermal_cycle_P_8x16}. 
While the range of $\beta$ indicative of metastability is shifted and 
reduced, we still observe a clear 
hysteresis loop at $\beta=4.95$ and $5.0$. 

For the two $\beta$ values we show the number of
BiCGStab iterations $N_{\mathrm{CG}}$ required for the
fermion matrix inversion during the calculation of the
Hamiltonian (\ref{eq:hamiltonian}) with a given stopping
criterion $10^{-14}$  in Figure~\ref{fig:NCG_thermal_cycle_P_8x16}.
We find a large gap in $N_{\mathrm{CG}}$ at the point of
metastability signals.
In the disordered phase $N_{\mathrm{CG}}$ is relatively small, 
whereas in the ordered phase $N_{\mathrm{CG}}$ is about a factor 2--3
larger.  
As we discuss in the next subsection the quark mass
is heavy at the point of metastability.  Indeed   
we find $m_{\mathrm{PS}}/m_{\mathrm{V}}\simeq$ 0.83--0.87.
Therefore the large value of $N_{\mathrm{CG}}$, which implies a small 
condition number of the Wilson-Dirac operator, in the ordered phase 
is not attributed to a physically small quark mass.

We suspect the metastability and gap to continue toward strong 
couplings below $\beta=4.9$.  We cannot confirm it, however, since 
the thermal cycles encounter a large Hamiltonian difference $dH>100$, 
and hence are turned back, before finding signals of the ordered phase.
For example, at $\beta$ = 4.8 this occurs at $\kappa=0.137$.
At this point $N_{\mathrm{CG}}= 55.4$, which is still a
relatively small number as seen in Figure~\ref{fig:NCG_thermal_cycle_P_8x16}.
The same is true for $\beta$ = 4.6 and $\kappa=0.134$, for
which $N_{\mathrm{CG}}=35.6$. 
Simply reducing the molecular dynamics step size or increasing the 
polynomial order does not resolve the occurrence of a large value of 
$dH$. Further studies are needed to understand the possible continuation of 
the ordered phase toward strong couplings.

\begin{figure}[tbp]
  \centering
  \includegraphics[scale=\figscale]{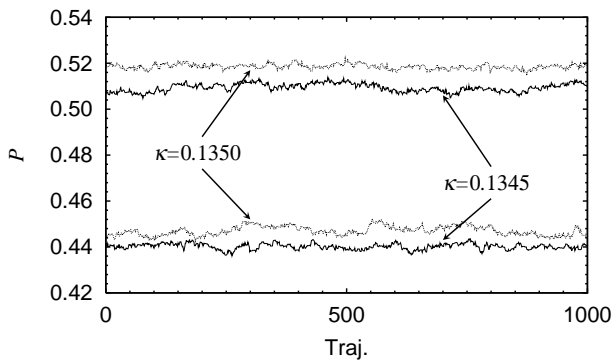}
  \caption{
    A typical example of the two-state signal on a $12^3
    \times 32$ lattice at $\beta=4.88$ and
    $c_{\mathrm{SW}}=2.15$.
    The plaquette history is shown.
    500--1000 trajectories are devoted to the thermalization from
    ordered/disordered configurations.
  }
  \label{fig:two-state_signal}
\end{figure}

\subsection{Exact simulation on a larger lattice}

To establish the nature of the hysteresis observed in
Figure~\ref{fig:thermal_cycle_P_8x16}, we perform
exact simulations starting from both ordered and disordered
configuration with fixed values of $\beta$ and $\kappa$ on
a $12^3 \times 32$ lattice. 
The ordered configurations are made at a larger $\kappa$
value, and the disordered configurations are generated in
the quenched limit $\kappa=0$ at the same $\beta$ value.

Simulation parameters, including $\beta$, $c_{\mathrm{SW}}$,
$\kappa$, the number of molecular dynamics steps
$N_{\mathrm{MD}}$, and the order of the polynomial $2n$,
are summarized in Table~\ref{tab:tab1}.
We remark that a polynomial order of only $2n \sim$ 30--200 are
necessary 
in order to achieve a $\sim$ 90\% acceptance rate 
$\langle P_{\mathrm{corr}}\rangle$ for the noisy Metropolis
test.

Figure~\ref{fig:two-state_signal} shows a representative result 
which demonstrates clear two-state signals 
persisting over 1000 trajectories.  
We confirm that the hysteresis seen in rapid thermal
cycles are not an artifact of our inexact simulations in which 
the HMC and global Metropolis tests are skipped.
Similar two-state signals are observed
using the $R$-algorithm on a $8^3\times 16$ lattice.
Our observation strongly suggests the existence of a
first-order phase transition separating the ordered and
disordered phases. 

\begin{table*}[t]
  \centering
  \caption{
    Lattice parameters used for the $12^3\times 32$
    simulations.
    $\langle P_{\mathrm{acc}}\rangle$ : 
    HMC acceptance rate, 
    $N_{\mathrm{MD}}$ : the number
    of MD step to proceed a unit trajectory.
    The molecular dynamics step size is given by
    $d\tau=1/N_{\mathrm{MD}}$.
    $\langle P_{\mathrm{corr}}\rangle$ :
    Global Metropolis test acceptance rate for
    correction factor, $2n$ : 
    the order of the polynomial. 
    The symbols in the `phase' column denote :
    L) smaller plaquette value (disordered phase), 
    H) larger  plaquette value (ordered phase), 
    M) signals are much unstable and long autocorrelation
    over 500 trajectories is observed.
  }
  \label{tab:tab1}
  \begin{ruledtabular}
  \begin{tabular}{cccccccccc}
   $\beta$ & $c_{\mathrm{SW}}$ & $\kappa$ &
      $\langle P_{\mathrm{acc}}\rangle [N_{\mathrm{MD}}]$ &
      $\langle P_{\mathrm{corr}}\rangle [2n]$ & phase &
      $am_{\mathrm{PS}}$ &$am_{\mathrm{V}}$ &
      $m_{\mathrm{PS}}/m_{\mathrm{V}}$ & $a_{r_{0}}^{-1}$ [GeV] \\ \hline
 4.88& 2.15& 0.1345 & 0.72[50]& 0.98[30] & L&         &         &     &         \\ 
     &     &        & 0.76[64]& 0.99[140]& H&         &         &     &         \\ \cline{3-10}
     &     & 0.1350 & 0.91[80]& 0.99[42] & L&         &         &     & 0.81(7) \\
     &     &        & 0.81[80]& 0.95[300]& H&         &         &     & 2.46(7) \\ \hline
 4.90& 2.14& 0.1340 & 0.70[50]& 0.90[24] & L&1.308(7) &1.555(14)& 0.84& 0.84(1) \\
     &     &        & 0.71[50]& 0.99[100]& H&0.682(14)&0.822(17)& 0.83& 1.381(3)\\ \cline{3-10}
     &     & 0.1343 & 0.73[50]& 0.86[24] & L&1.247(5) &1.481(21)& 0.84& 0.83(1) \\
     &     &        & 0.78[64]& 0.98[140]& H&0.458(10)&0.594(21)& 0.77& 1.90(4) \\ \cline{3-10}
     &     & 0.1345 & 0.73[50]& 0.96[36] & L&1.185(11)&1.405(14)& 0.84& 0.79(1) \\
     &     &        & 0.74[64]& 0.98[200]& H&0.433(14)&0.587(18)& 0.74& 1.99(4) \\ \cline{3-10}
     &     & 0.1346 & 0.85[80]& 0.80[120]& H&0.439(14)&0.559(24)& 0.78& 2.44(7) \\ \hline
 4.95& 2.11& 0.1325 & 0.76[50]& 0.98[30] & L&         &         &     &         \\ \cline{3-10}
     &     & 0.1328 & 0.72[50]& 0.97[30] & L&1.285(9) &1.496(16)& 0.86& 0.827(4)\\
     &     &        & 0.74[50]& 0.99[70] & H&0.844(13)&0.970(16)& 0.87& 1.39(1) \\ \cline{3-10}
     &     & 0.1330 & 0.73[50]& 0.98[70] & H&         &         &     &         \\ \hline
 4.97& 2.10& 0.1320 & 0.72[50]& 0.98[34] & L&         &         &     &         \\ \cline{3-10}
     &     & 0.1322 & 0.77[50]& 0.96[30] & L&         &         &     & 0.85(1) \\ \cline{3-10}
     &     & 0.1323 & 0.77[50]& 0.99[60] & H&         &         &     & 1.35(2) \\ \cline{3-10}
     &     & 0.1325 & 0.73[50]& 0.99[60] & H&         &         &     &         \\ \hline
 5.00& 2.08& 0.1310 & 0.76[50]& 0.95[24] & L&         &         &     &         \\ \cline{3-10}
     &     & 0.1313 & 0.75[50]& 0.89[26] & L&         &         &     &         \\ \cline{3-10}
     &     & 0.1314 &         &          & M&         &         &     &         \\ \cline{3-10}
     &     & 0.1315 &         &          & M&         &         &     &         \\ \cline{3-10}
     &     & 0.1320 & 0.85[64]& 0.99[60] & H&         &         &     & 1.53(2) \\ \cline{3-10} 
     &     & 0.1330 &0.90[100]& 0.99[100]& H&0.583(5) &0.692(6) & 0.84& 2.06(6) \\ \cline{3-10} 
     &     & 0.1338 & 0.90[80]& 0.93[100]& H&0.475(10)&0.569(9) & 0.83& 2.58(4)
  \end{tabular}
  \end{ruledtabular}
\end{table*}

In Table~\ref{tab:tab1} we list results for 
pseudo-scalar meson mass $m_{\mathrm{PS}}$ and vector meson mass  $m_{\mathrm{V}}$ in 
lattice units.  We have also calculated the Sommer scale $r_0$
using the condition $r_0^2 dV(r)/dr|_{r=r_0}=1.65$
on the static potential.  The lattice spacing estimated from 
the phenomenological value $r_0$ = 0.49~fm is listed in Table~\ref{tab:tab1}.

\begin{figure}[tbp]
  \centering
  \includegraphics[scale=\figscale]{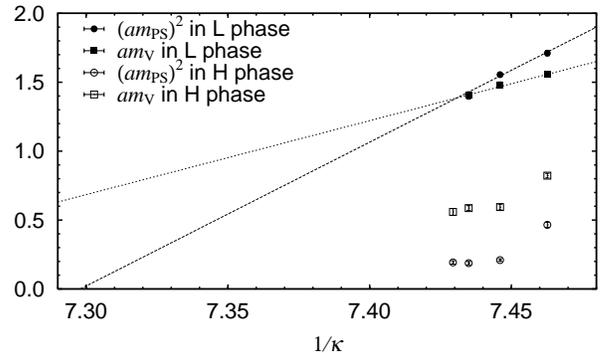}
  \caption{
    Chiral extrapolation of $(am_{\mathrm{PS}})^2$ and $am_{\mathrm{V}}$
    at $\beta=4.9$ on a $12^3\times 32$ lattice. In `H' phase we observe
    larger plaquette values (ordered phase), and in `L' phase smaller plaquette 
    values are observed (disordered phase).
  }
  \label{fig:chiral_b4.9}
\end{figure}

In Figure~\ref{fig:chiral_b4.9} we plot $(am_{\mathrm{PS}})^2$ and $am_{\mathrm{V}}$
at $\beta=4.9$ as functions of $1/\kappa$.  If we extrapolate 
the data in the disordered phase (solid symbols) linearly as is conventionally 
made, we find $\kappa_c=0.13703(29)$ and $m_{\mathrm{V}}(\kappa_c)=0.67(11)$ which 
translates into $a^{-1}_{m_{\rho}}=0.87(14)$~GeV as a rough estimate of lattice spacing.  
These are quite natural values comparable to those encountered in quenched 
and two-flavor simulations.  However, before reaching this point, 
the three-flavor system makes a transition into 
an ordered phase in which hadron masses are drastically reduced. 

If one looks at the lattice spacing determined from $r_0$ in 
Table~\ref{tab:tab1}, we see that $a^{-1}_{r_0}\approx 0.8$~GeV in the disordered 
phase, which is consistent with the spectrum estimate from $m_{\rho}$ above, 
while $a^{-1}_{r_0}\approx 2$~GeV in the ordered phase is much larger.  
Furthermore pion mass squared does not seem to decrease 
in the ordered phase.  We then suspect that physical results cannot be obtained 
with simulations in the ordered phase.

\subsection{Phase diagram}

\begin{figure}[tbp]
  \centering
  \includegraphics[scale=\figscale]{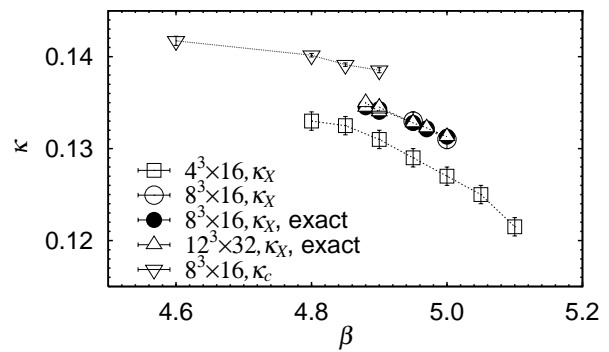}
  \caption{
    Phase diagram in the $(\beta,\kappa)$ plane.
    Data for the phase transition points $\kappa_X$ are from
    the rapid thermal cycles on  
    $4^3\times 16$ (open squares) and
    $8^3\times 16$ (open circles) lattices.
    Filled circles and open up triangles are
    from the exact simulation on 
    $8^3\times 16$ and
    $12^3\times 32$ lattices, respectively.
    The critical hopping parameter $\kappa_c$ (open down triangles) 
    is estimated
    using an extrapolation from smaller $\kappa$ values.
  }
  \label{fig:phase_diagram}
\end{figure}

In Figure~\ref{fig:phase_diagram} we plot the location of
the observed phase transition $\kappa_{X}(\beta)$ in the
($\beta,\kappa$) plane for various lattice sizes.  
Open squares for a $4^3 \times 16$ lattice and open circles for a 
$8^3\times 16$ lattice show the point of hysteresis observed with the inexact 
algorithm, while filled circles and open up triangles correspond to 
the point where two-state 
signals such as in Figure~\ref{fig:two-state_signal} are observed with the 
exact algorithm.

\begin{figure}[tbp]
  \centering
  \includegraphics[scale=\figscale]{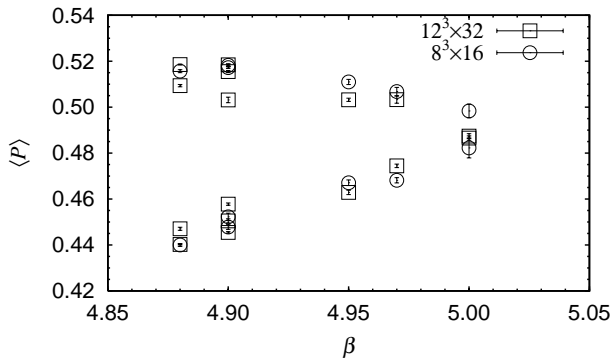}
  \caption{
    Plaquette expectation values in the two phases on the
    first-order phase transition line.
    Data from the $8^3\times 16$ and $12^3\times 32$
    lattices are plotted as a function of $\beta$.
  }
  \label{fig:gap}
\end{figure}

The location of the phase transition line significantly
moves when we increase the lattice size from $4^3 \times 16$
to $8^3 \times 16$. 
However, it stays at the same place when
the lattice size is further increased to 
$12^3\times 32$, which strongly suggests that the
first-order transition line persists in the infinite volume
limit at zero temperature.  
In fact, the gap in the value of $\langle P\rangle$ does not 
significantly change from $8^3\times 16$ to $12^3\times 32$,
as shown in Figure~\ref{fig:gap}.

Figure~\ref{fig:gap} also shows that the gap in the
plaquette expectation value decreases toward larger $\beta$
and vanishes around $\beta=5.0$.
Since we observe no sign of second-order transition at
weaker couplings (Figure~\ref{fig:thermal_cycle_P_8x16}),
the transition appears to terminate at $\beta\simeq 5.0$.

Toward the strong coupling regime, we expect the phase transition line 
to continue below the last point of $\kappa_{X}$ at $\beta=4.88$ in 
Figure~\ref{fig:phase_diagram}.  Indications are that 
the gap tends to become large toward lower $\beta$ values 
(see Figure~\ref{fig:gap}), while the pseudo-scalar-to-vector meson mass 
ratio on the gap is almost independent of $\beta$.

The critical hopping parameter $\kappa_c$ drawn by downward 
triangles in Figure~\ref{fig:phase_diagram} represents a rough estimate 
based on the number of iterations of the BiCGStab solver,
$N_{\mathrm{CG}}$, in the calculation of 
Hamiltonian (Eq.~(\ref{eq:hamiltonian}))
on a $8^3\times 16$ lattice.
$N_{\mathrm{CG}}$ is sampled during the thermal cycles
and the estimates are obtained by linearly extrapolating
$1/N_{\mathrm{CG}}^2$ as a function of $1/\kappa$. 
As mentioned earlier the thermal cycles do not extend far 
toward large values of $\kappa$ at the $\beta$ values below
4.88. Therefore the estimate of $\kappa_c$ may have a large
uncertainty.  
We also emphasize whether there actually exists the critical 
$\kappa_c$ where the pion mass vanishes in the ordered phase 
is an open issue.  In fact the pion mass measured in the ordered phase 
at $\beta=4.9$ shown in Figure~\ref{fig:chiral_b4.9} is almost constant 
as a function of $\kappa$.  

\subsection{Practical implications}

Our findings expose a serious practical problem on simulations 
using the $O(a)$-improved Wilson fermion action in combination with 
the plaquette gauge action.  
To guarantee a smooth extrapolation to the continuum limit, 
we should carry out simulations at coupling weaker than the 
termination point of the first-order transition at $\beta\approx 5.0$. 
As shown in Table~\ref{tab:tab1}, the lattice spacing 
estimated from $r_0=0.49$~fm at $\beta$ = 5.0 is $1/a$ = 
1.53(2)~GeV ($\kappa$ = 0.1320), 
2.06(6)~GeV ($\kappa$ = 0.1330) and 
2.58(4)~GeV ($\kappa$ = 0.1338).
The largest $\kappa$ value ($\kappa=0.1338$) still
corresponds to a heavy quark 
($m_{\mathrm{PS}}/m_{\mathrm{V}}\sim 0.83$).
Taking the significant $\kappa$ dependence of $1/a$ into
account, the lattice spacing in the chiral limit would be even
larger, possibly greater than 3~GeV. 
A 2~fm lattice would then require a 
$30^3\times 60$ volume or larger.  
Large-scale simulations starting at such fine lattices and large 
lattice volumes are too  computer time consuming even with 
high-end supercomputers available at present.

\section{Phase structure for improved gauge actions}
\label{sec:improved_actions}

If the first-order transition observed for the plaquette gauge action 
is a lattice artifact, one may expect that it can be eliminated by 
improving the gauge action since scaling
toward the continuum limit is much improved and the lattice
artifacts are expected to be suppressed for these actions. 

\begin{figure}[tbp]
  \centering
  \includegraphics[scale=\figscale]{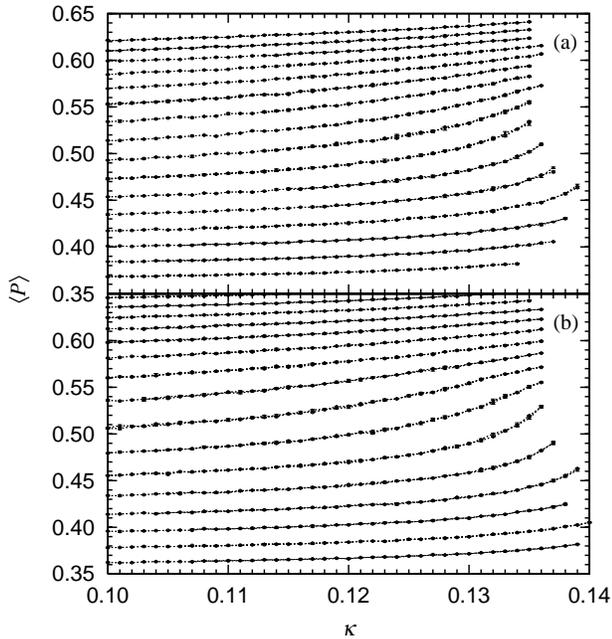}
  \caption{
    Thermal cycles of plaquette expectation value $\langle
    P\rangle$ on a $8^3\times 16$ lattice,
    (a) with the RG improved gauge action at
    $\beta=1.50$--$2.25$ in steps of 0.05, 
    and (b) with tadpole improved Symanzik gauge action at
    $\beta=3.20$--$4.70$ in steps of 0.10 
    (from bottom to top).
  }
  \label{fig:thermal_cycle_I_8x16}
\end{figure}

Here we test two types of improved gauge actions, both
of which are defined with (\ref{eq:gluon_actions}): 
one is the RG improved action~\cite{Iwasaki:1983ck} and the
other is the $O(a^2)$-improved (L\"uscher-Weisz (LW)) 
action~\cite{Luscher:1984xn}. 
In the LW case, the tadpole factor $\langle P\rangle$ in $c_1$ is 
self-consistently determined at $\kappa=0$ for each $\beta$.
The clover term is also determined by tadpole-improved
one-loop perturbation theory (Eq.~(\ref{eq:c_sw_plaqutte})) with    
$\langle P\rangle$ at $\kappa=0$.
Figure~\ref{fig:thermal_cycle_I_8x16} shows the results of
the thermal cycles on a $8^3 \times 16$ lattice.
The simulation conditions such as the values for $2n$ and $d\tau$, 
and skipping of the HMC and global Metropolis tests, \textit{etc.}
are the same as those with the unimproved gauge action.
In contrast to the case of the plaquette gauge action,
we do not observe any remnant of hysteresis loop.
This indicates that there is no first-order phase transition with these
improved actions.

\section{Possible origin of the first-order transition}
\label{sec:Discussions}
\subsection{Bulk phase transition in the fundamental-adjoint coupling plane}

In the pure SU(3) lattice gauge theory having both fundamental 
and adjoint couplings $\beta$ and $\beta_A$, 
a bulk first-order phase transition exists in the strong coupling 
regime~\cite{Greensite:hw,Bhanot:1981pj,Heller:1995bz}.
The transition line starts at the purely adjoint point
$(\beta,\beta_A)=(0,6.5(3))$ and extends toward larger $\beta$ with
decreasing $\beta_A$. It terminates at 
$(4.00(7),2.06(8))$ and never crosses the purely fundamental
line $\beta_A=0$, so that the pure gauge lattice theory with
the fundamental representation is smoothly connected to the
continuum limit $\beta=\infty$.
Near the critical endpoint the correlation function of the
$0^{++}$ glueball channel diverges, but the scaling of other
observables is not much affected~\cite{Heller:1995bz}.

This phase transition shares many
properties with the one we found in three-flavor lattice QCD. 
The first-order transition separates ordered and disordered
phases (large and small plaquette expectation values,
respectively), and their lattice spacing measured through
the string tension is largely different.
Furthermore, the transition has a bulk nature,
\textit{i.e.} it remains in the infinite volume limit.

One may suspect that dynamical Wilson fermions effectively induce the adjoint
gauge coupling, which gives rise to the bulk transition.
This possibility was explored for two-flavor unimproved
Wilson fermion action in Ref.~\cite{Blum:1994xb}.
They measured the strength of the induced adjoint coupling on
the dynamical configurations and found that it is slightly
negative, as opposed to the expectation. 
The clover term might enhance the induced adjoint coupling
as it has a $1\times 1$ Wilson loop structure, but it is to
be confirmed either analytically or numerically.

\subsection{Breakdown of parity-flavor symmetry}

For two-flavor lattice QCD with the Wilson-type quark action, Sharpe and 
Singleton~\cite{Sharpe:1998xm} 
carried out an analysis of parity-flavor symmetry 
using chiral Lagrangian techniques.  They pointed out that, depending on 
the sign of an $O(a^2)$ term, there may be a line of first-order phase transition 
along which pion mass is non-zero and the chiral condensate flips sign, rather 
than a pair of second-order transition lines along which pion mass vanishes and 
parity-flavor symmetry breaks down. 

It is straightforward but complicated to extend this type of analysis to 
the three flavor case.  
Since there are three $O(a^2)$ terms allowed in the chiral Lagrangian, 
rather than a single term for the two flavor case, 
predictions are less definite.  Nonetheless one may
similarly expect that first-order transitions may occur depending on 
the coupling of the three terms. 

Singular phenomena have been observed with dynamical
Wilson-type fermion simulations in a variety of contexts, and our finding 
is one more of the list of such phenomena.  
While it is not clear at present if the above analysis offer an understanding of 
these phenomena, we attempt to discuss them for orientation of future studies.

\subsubsection{Two flavor case}

Farchioni {\it et al}.~\cite{Farchioni:2004us} recently reported 
a first-order phase transition for the plaquette gauge action and unimproved 
Wilson quark action at $\beta=5.2$ and $\kappa\approx 0.1715$.  They 
suggested that this phase transition may be understood within 
the Sharpe-Singleton analysis. 

In an old work on the finite temperature phase transition of two-flavor QCD 
with the Wilson fermion action, an unexpected strong first-order transition 
was found~\cite{Blum:1994eh}.  It was suggested that this is a bulk 
phase transition since the Polyakov loop does not jump at the transition 
point below $\beta\simeq$ 5.0 while the plaquette expectation value shows 
strong metastability.  

It is plausible that the two findings refer to the same bulk first-order 
phase transition.  Numerically, Ref.~\cite{Blum:1994eh} found a metastability at 
$\beta=5.22$ and $\kappa=0.17$, in a close proximity of that in 
Ref.~\cite{Farchioni:2004us}, and the plaquette values in the two phases 
reported by the two studies are in agreement.  

It is not clear if such the first-order phase transition persists when 
the Wilson quark action is improved by the addition of the clover 
term.  For non-perturbatively $O(a)$-improved Wilson fermion 
action~\cite{Jansen:1998mx}, most physical observables, such as hadron
masses and matrix elements, measured in the past 
simulations~\cite{Allton:2001sk,Aoki:2002uc} do not show singular behavior. 
On the other hand, the mass of $0^{++}$ glueball is surprisingly lower 
than in the quenched case~\cite{Hart:2001fp}, perhaps hinting at the 
presence of a nearby singularity in the coupling constant space. 
Also the lattice artifact in the measurement of the light  
quark mass through the axial-Ward-Takahashi identity is found to be rather 
large~\cite{Sommer:2003ne}.

We also note that the strong first-order transition for the unimproved 
Wilson quark action disappears if the gauge action is 
improved~\cite{Bernard:1997an,Iwasaki:1996ya}. 

\subsubsection{Three-flavor case}

The report by Farchioni {\it et al.} of a first-order phase transition 
for the unimproved Wilson quark action and the plaquette gauge action 
raises the possibility that a similar first-order transition may be present 
for the three flavor case.  Indeed, in previous finite-temperature studies, 
a large lattice artifact was found for this action 
combination~\cite{Iwasaki:1996zt}: at the point of finite-temperature 
transition, the light quark mass measured from the axial-Ward-Takahashi
identity jumps for $\beta\lesssim$ 5,  contrary to the expectation that 
the Ward-Takahashi identity holds at any physical phases with an identical  
value for the measured quark mass for the same bare parameters. 

\begin{figure}[tbp]
  \centering
  \includegraphics[scale=\figscale]{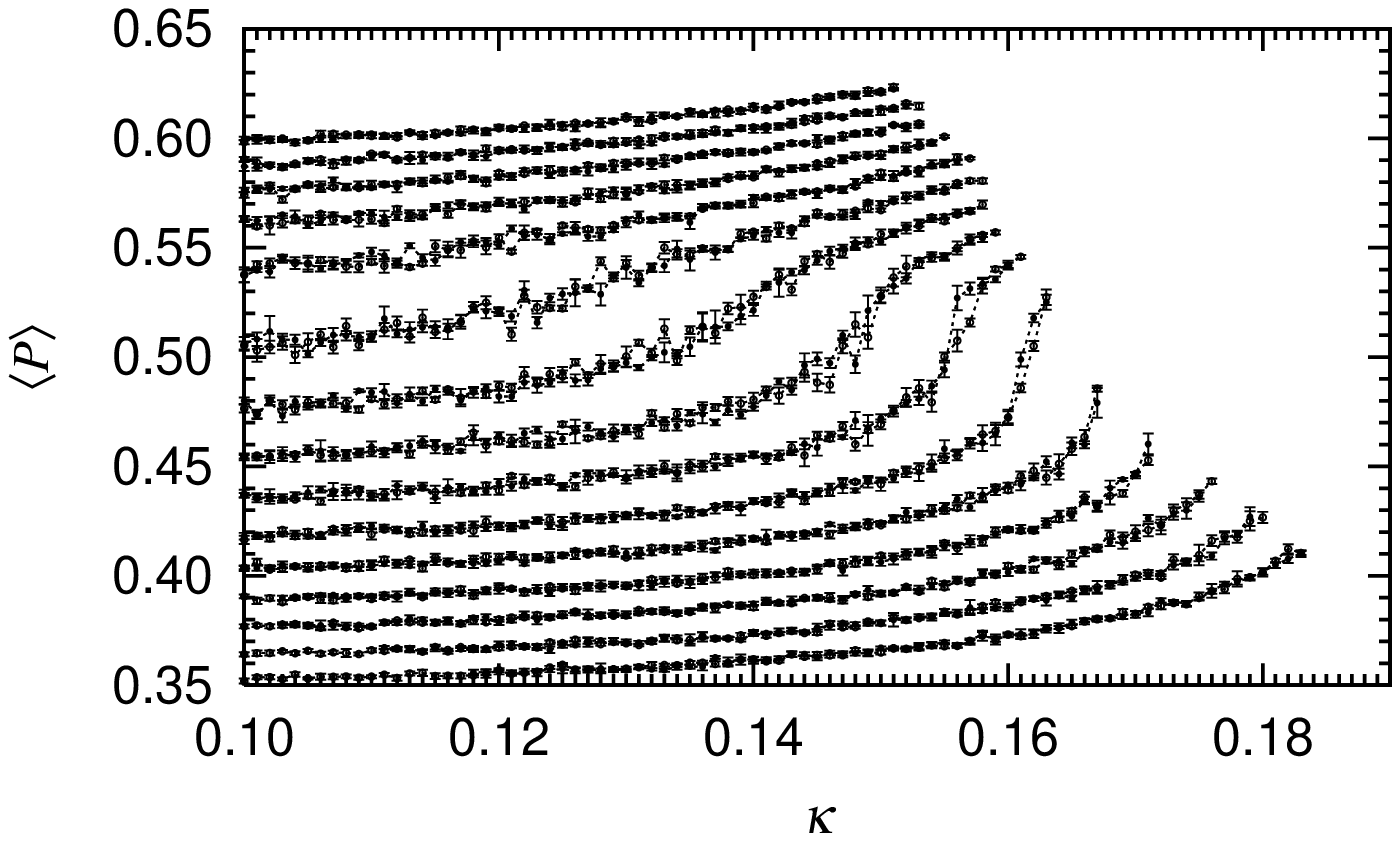}
  \\
  \includegraphics[scale=\figscale]{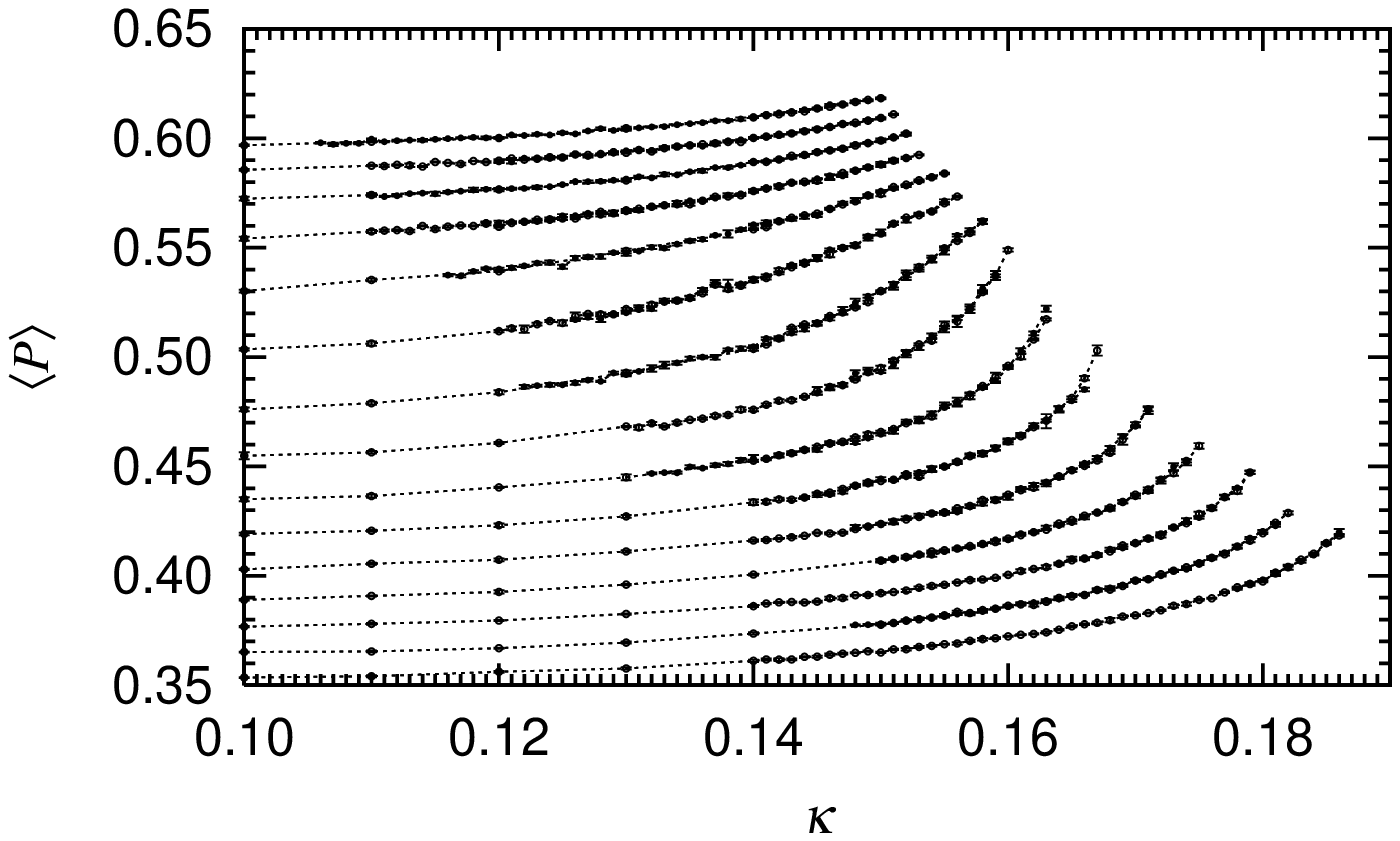}
  \caption{
    Thermal cycles of the plaquette expectation value
    $\langle P\rangle$ for
    the plaquette gauge action and
    unimproved Wilson fermion on $4^3\times 16$ (top) and 
    on $8^3\times 16$ (bottom) lattices.
    The $\beta$ values in the region 4.6--6.0 are scanned.
    Lines show different $\beta$ values in steps of 0.1
    (from bottom to top).
  }
  \label{fig:thermal_cycle_PW}
\end{figure}

We have attempted an initial thermal cycle study on $4^3\times
16$ and $8^3\times 16$  lattices with unimproved Wilson quark action,
and the results are shown in Fig.~\ref{fig:thermal_cycle_PW} for
the $\beta$ values in the range 4.6--6.0.
We observe a signature of phase transition on the 
$4^3\times 16$ lattice (top) at $\beta=5.1--5.2$, while metastabilities  
are not apparent on the $8^3\times 16$ lattice (bottom).

\begin{figure}[tbp]
  \centering
  \includegraphics[scale=\figscale]{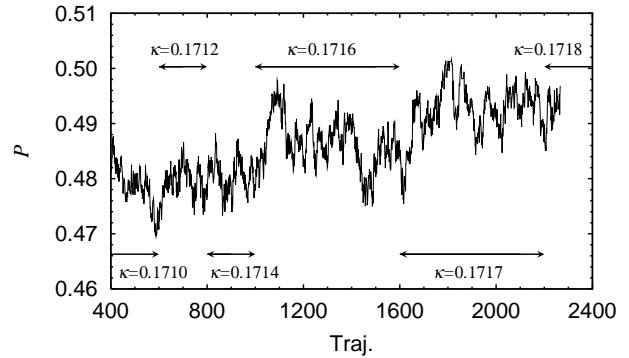}
  \caption{
    Plaquette history of the $N_f=3$ exact simulation with
    the plaquette gauge action and the unimproved Wilson
    fermion action at $\beta$ = 5.0.
    The hopping parameter $\kappa$ is changed from 0.1710 to
    0.1718 as indicated in the plot.
  }
  \label{fig:plaqhis}
\end{figure}

The absence of the gap on the $8^3\times 16$ lattice does not
necessarily mean that the chiral and continuum limit can be
smoothly reached, since the thermal cycle above 
does not cover the small quark mass region.
In order to cover this region we carry out a dedicated
run at $\beta$ = 5.0. 
Figure~\ref{fig:plaqhis} shows the  history of plaquette over
the HMC trajectories starting from $\kappa=0.1710$,  
where the thermal cycle ended, up to 0.1718.
We observe large fluctuations of plaquette at $\kappa$ =
0.1716 and 0.1718, which my be hinting at the possible presence of 
a phase transition. 

Clearly further work is needed to reach a comprehensive understanding 
of the phase structure of lattice QCD with Wilson-type fermion action 
for two and three flavor cases.

\section{Conclusions}
\label{sec:Conclusions}

We have reported the existence of an unexpected phase transition in
three-flavor lattice QCD with the $O(a)$-improved Wilson
fermion action.
It appears in the strong coupling regime $\beta\lesssim$ 5.0
if one uses the standard Wilson plaquette gauge action, while there is no
indication of such a phase transition for improved gauge actions.
The phase transition persists for large lattice volumes,
and is likely a bulk phase transition. 

Our findings pose a serious practical problem on simulations 
using the $O(a)$-improved Wilson fermion action in combination with 
the plaquette gauge action.  To avoid unphysical effects of the bulk 
transition, one has to carry out simulations at couplings weaker than its  
end point, but the lattice spacing is already smaller than 
$a\sim 0.1$~fm there, necessitating large lattice volumes and 
hence large computing resources. 

This circumstance motivates us to employ 
the RG-improved gauge action, for which we observe no strange phase structure, 
for large-scale three-flavor simulations. 
A non-perturbative determination of the improvement
coefficient $c_{\mathrm{SW}}$ for a full O(a) improvement 
has been made using the Schr\"{o}dinger functional 
method~\cite{Aoki:2002vh,Ishikawa:2003ri,Yamada:2004ja}, and  
preliminary results on the light hadron spectrum have
already been presented in Ref.~\cite{Kaneko:2003re}.

Finally, the recent report that the two-flavor system with the unimproved 
Wilson action also has a first-order transition, and that it may be 
understood with the context of the Sharpe-Singleton analysis on realizations 
of the parity-flavor broken phase raises an interesting problem that we need 
to clarify for phenomenological applications of full QCD simulations with 
Wilson-type quark actions.

\begin{acknowledgments}
  This work is supported by the Supercomputer Project No.~98
  (FY2003) of High Energy Accelerator Research Organization
  (KEK), and also in part by the 
  Grant-in-Aid of the Ministry of Education (Nos. 10640246,
  11640294, 12014202, 12640279, 12740133, 13135204, 13640260, 13740169,
  14046202, 14740173, 15540251, 15540279
  and 16028201.
\end{acknowledgments}

\end{document}